# Participating in a Computer Science Linked-courses Learning Community Reduces Isolation

Amber Settle, James Doyle, Theresa Steinbach
DePaul University
243 S. Wabash Avenue
Chicago, IL 60604
(312) 362-8381
asettle@cdm.depaul.edu, jdoyle12@mail.depaul.edu, tsteinbach@cdm.depaul.edu

## ABSTRACT
In our previous work we reported on a linked-courses learning community for underrepresented groups in computer science, finding differences in attitudes and resource utilization between students in the community and other programming students. Here we present the first statistically significant differences in pre- to post-quarter student attitudes between those in the learning community and others taking equivalent programming classes. We find that students in the learning community are less likely to feel isolated post-quarter than other programming students. We also present results showing differences in resource utilization by learning-community participants.

## 1. INTRODUCTION
In an effort to improve retention of underrepresented students at our institution, a linked-courses learning community was created beginning in the first term of the academic year 2014-2015 and continued in the first term of the 2015-2016 academic year. Entering first-year students majoring in areas that require an introductory Python programming course were also enrolled in a required first-year general education class focused on the digital divide. The students recruited for the community were members of underrepresented groups in computing, including both women and non-white men.

While the ultimate goal is improved retention, which by definition requires time to measure, attitude changes can provide an early indication of the potential impact, or lack of impact, of the learning community. To gain insight we previously examined attitude differences between students participating in the learning community and students enrolled in the Python programming course but not in the learning community [6, 7, 8]. In this report we combine the data from the first two cohorts of the learning community to find the first significant differences in pre- to post-quarter student attitudes between students in the learning community and general programming students. We find that students in the learning community are less likely to feel isolated post-quarter than general programming students, confirming the result seen in the learning community alone [8]. This result is promising given that isolation is an issue for underrepresented groups [5].

## 2. BACKGROUND
The work in this paper examines differences in student attitudes and resource utilization between students participating in our linked-courses learning community and equivalent populations in our college. In this section we provide background information about the courses and activities in the community, the survey instrument used to measure student attitudes and resource use, and results from analysis of earlier data sets.

### 2.1 Courses and Activities
The learning community was created for first-quarter freshman who are either a man of color or a woman of any race in majors that require Introduction to Computer Science I (CS1), an introductory programming course taught in Python that focuses on problem solving. The included majors are computer science, math and computer science, computer game development in the gameplay and systems concentrations, and cybersecurity. Our college uses several interventions recommended in the literature in the introductory programming sequence for majors, including closed labs with collaborative activities, differentiated courses for novice and experienced programmers, and engaged and enthusiastic faculty [1, 2, 3, 4]. It should be noted that our institution is large, with approximately 2000 undergraduates in our college. While students in any of these majors have a fixed set of classes to take during their first year including the CS1 course, there are many different sections from which to choose. For example, there are typically eight sections of the CS1 course offered during the first quarter of each academic year. One of the sections of the CS1 course is reserved for learning community students.

Every freshman at our institution is required to take a Chicago Quarter class. These classes are designed to acquaint first-year students with our institution and the metropolitan community, neighborhoods, cultures, institutions, organizations, and people of Chicago. The courses also have a "Common Hour," which addresses issues of transition for first-year students, including academic success skills and educational and career planning. Students participating in the linked-courses learning community were enrolled in a section of an Explore Chicago class focusing on the digital divide and specifically on the social issues surrounding access to information and communications technology (ICT).

To enhance the sense of community among students, the CS1 and Explore Chicago courses are scheduled in back-to-back slots. Immediately following the CS1 sessions are the Explore Chicago class sessions, and students typically walk together between the two classes. With an institution as large as ours, having the opportunity as first-year students to share classes and walk together between those classes is unique to the learning community.

In addition to these two classes, students in the learning community participated in a variety of co-curricular and extra-curricular activities, including an open house, study sessions for the midterm and final exams, a midterm gaming celebration, employer visits, and tours of specialized facilities available in our college. The open house and gaming sessions are held at the instructors' homes, with the intention of bringing the participants closer to the instructors as well as to each other. There were two cohorts of the learning community, the first during the first quarter of the 2014-2015 academic year and the second during the first quarter of the 2015-2016 academic year. Activities for the first cohort were for the most part optional, and activities for the second cohort were either mandatory or provided extra credit for attendees. As a result, the second cohort had improved participation in the co-curricular and extra-curricular activities [8].

Finally we note that the instructor for the CS1 course and the Explore Chicago course were the same for both the first and second cohorts.

## 2.2 Survey Instrument

The goal of the linked-courses learning community is to engage the students in an effort to improve their feelings of belonging and confidence, improve their study habits, and ultimately, improve their retention in the courses and degree program. While improved retention is the goal measuring the potential impact of the community early is beneficial for a variety of reasons.

As an early measure of whether students experienced a change in attitudes and habits by taking part in the learning community we developed and administered a survey. The survey has 33 questions to measure attitudes toward computing and programming. Each of the attitude questions were answered using a five-point Likert scale of 5 = strongly agree, 4 = agree, 3 = neutral, 2 = disagree, and 1 = strongly disagree. There was one additional question that asked students about their utilization of resources for study. The questions in the survey are provided below:

1. I plan to major in a technology-related degree
2. I am sure that I can learn programming
3. I am sure I can do advanced work in computer science
4. I think I could handle more difficult programming problems
5. I can get good grades in computer science
6. I have a lot of self-confidence when it comes to programming
7. I'm not good at programming
8. For some reason even though I work hard at it, programming seems unusually hard for me
9. Computer science has been my worst subject
10. It would make me happy to be recognized as an excellent student in computer science
11. I'd be happy to get top grades in computer science
12. If I got good grades in computer science, I would try to hide it
13. I'll need programming for my future work
14. Knowing programming will help me earn a living
15. I will use programming in many ways throughout my life
16. Taking computer science courses is a waste of time
17. Once I start trying to work on a program, I find it hard to stop
18. I am challenged by programming problems I can't understand immediately
19. I am easily frustrated by difficult programming problems
20. I do as little work in computer science courses as possible
21. I like talking with my friends about programming
22. I like to program in my spare time
23. I belong in the computing field
24. I feel isolated in computer science courses
25. I am part of a community of programmers
26. Computer science offers good financial opportunities after graduation
27. Computer science allows me to be creative
28. Computing offers diverse and broad opportunities
29. I have a lot of support that will help me to succeed in computer science courses
30. Computer science provides opportunities to make a difference in the world
31. I have a lot of friends who are interested in computing
32. My family is happy that I am taking computer science courses
33. I have had good teachers in my computer science courses

There was one additional question that asked students: "Outside of your classroom studies, what are your resources for learning/obtaining new computing skills? Mark all that you use and rank them in order of use. The one you use most should be ranked as 1, the one you use next often as 2, etc." The resources listed were: Friends/peers, Internet/web sites, professional organizations, self study, family members, tutors, faculty, and other (please specify). Students were instructed to leave a resource blank if they did not use it.

The survey was administered pre-quarter and post-quarter in all CS1 Python classes during the 2013-2014 academic year and during the first quarters of the 2014-2015 and 2015-2016 academic years.

## 3. EVALUATION

In our evaluation of the impact of the linked-courses learning community on students' attitudes about computing and utilization of resources, we compare the results of survey responses from students in the learning community with the responses from students enrolled in a section of the CS1 course but not participants in the community. We consider the data from both the 2014-2015 and 2015-2016 academic years, looking for both consistent and inconsistent impacts across the years.

## 3.1 Demographics

There were two basic populations considered in the study: students who took part in the learning community during the 2014-2015 or 2015-2016 academic year, and students who participated in the CS1 course during the first quarter of the 2014-2015 or 2015-2016 academic year but were not part of the learning community.

### 3.1.1 Included subpopulations

There were 332 students who responded to both the pre- and post-quarter surveys. Of these, 296 students took the CS1 course, and 36 took part in the learning community. The 2014-2015 academic year saw responses from 198 students in the CS1 group and 13 students in the learning community. In the following year, there were 98 responses from the CS1 course and 23 from the learning community.

The table below shows the post-quarter gender demographics. The pre-quarter demographics differed only in the 2015-2016 responses. In the pre-quarter responses one additional person marked "no response" and there was one less response in the male category.

**Table 1. Post-quarter gender for each population**

| Gender | 2014 CS 1 | | 2014 LC | | 2015 CS 1 | | 2015 LC | |
|---|---|---|---|---|---|---|---|---|
| | Ct. | % | Ct. | % | Ct. | % | Ct. | % |
| Male | 146 | 74% | 10 | 77% | 78 | 80% | 14 | 61% |
| Female | 52 | 26% | 3 | 23% | 17 | 17% | 9 | 39% |
| No response | 0 | 0% | 0 | 0% | 3 | 3% | 0 | 0% |
| Total | 198 | | 13 | | 98 | | 23 | |

### 3.1.2 Excluded subpopulations

In addition to the subpopulations above, data was also gathered as to the students' ethnicities, majors, and prior academic progress. These factors were excluded from the statistical model due to the lack of sufficient data to properly detect interaction effects with their inclusion. As additional data is gathered in future offerings of the learning community, we expect that it will become possible to analyze interactions with these factors.

## 3.2 Attitude Questions

The responses to the attitude questions were analyzed using a multi-way unbalanced ANOVA test to determine if any of the subpopulations showed statistically significant variation in their mean responses. The model examined basic and interaction effects between four categories: pre- vs. post-quarter results (time), 2014-2015 vs. 2015-2016 academic years (year), learning community vs. CS1 courses (coursetype), and the gender of the student (gender). The major, quarter, and ethnicity factors were not included in the interaction model, as there was not enough data to calculate least-squares means in many of the post-hoc tests with such a complex model.

When a basic or interaction effect was shown at the $p=.05$ or better significance level on the overall F-test, a post-hoc test of least-squares means with Tukey-Kremer adjustment for multiple comparisons was performed. If the results showed interaction between time and one or more other factors, the post-hoc analysis was done using a gainscore approach, calculating the change in response provided by the same student during the quarter.

This analysis focuses on those statistically significant results that showed an interaction effect between the time and coursetype classifications. Such an effect indicates that the learning community may have had an impact on the change in response from pre- to post-test.

The results of Q24: "I feel isolated in computer science courses," showed a statistically significant difference ($F(1,320)=6.29$, $p=.0127$) between the learning community and CS1 groups in the over-time change in the dependent variable (the gainscore). Both groups had similar mean responses prior to the quarter, but the learning community students showed a statistically significant drop between pre- and post- quarter responses while the CS1 students did not.

Table 2 shows the means for CS1 and learning community students before and after the course. Table 3 shows the change in response (the gainscore) for each group, as well as the result of the post-hoc t-test comparing the two. For this test, $H_0$ is that the mean gainscore of the two groups are equal.

**Table 2. Mean responses for LC and CS 1 groups**

| Q | CS 1 Pre | CS 1 Post | LC Pre | LC Post |
|---|---|---|---|---|
| 24 | 2.32 | 2.18 | 2.33 | 1.62 |

**Table 3. Post-hoc test for differences LC vs. CS 1**

| Q | CS 1 Gainscore | LC Gainscore | Result (T-K adj. p) |
|---|---|---|---|
| 24 | -0.13 | -0.71 | $t(320)=2.51, p=.0127$ |

The results of two questions showed statistically significant two-way interaction between the gainscores of learning community and CS1 groups and the two different academic years. These occurred on Q8: "For some reason, even though I work hard at it, programming seems unusually hard for me" ($F(1,320)=9.09$, $p=.0028$) and Q27: "Computer science allows me to be creative" ($F(1,320)=7.85$, $p=.0054$).

## 3.3 Resource Question

Responses to the resource question consisted of a relative ranking of each resource ranging from 1 to 8, with 1 being the most frequently used resource. Some responses contained multiple resources ranked in a tie. To properly account for ties and resources that were left off one but not both of the pre- and post-test, all responses were analyzed by converting the ordinal ranking for each resource provided by the student into a score reflective of its position relative to other resources. For each student's response, a resource was assigned a score calculated by adding the number of lower-ranked resources to half the number of equally-ranked resources. Thus, a higher score is reflective of a more-frequently-used resource than a lower score, and all scores were between 0 and 7.

The scores for each resource were then analyzed in a similar manner to the attitude questions. An overall multi-way unbalanced ANOVA test was used to determine which, if any, of the subpopulations showed a statistically significant variation on their mean responses. Basic and interaction effects were investigated using the same categories as for the attitude questions; again the major, quarter, and ethnicity factors were left out of the model due to data size limitations. The same gainscore approach was used in post-hoc testing effects that showed interaction with the time factor.

The analysis of resources "Internet/web sites" ($F(1,320)=8.47$, $p=.0039$) and "professional organizations" ($F(1,320)=9.84$, $p=.0019$) revealed statistically significant differences between the LC and CS1 groups in the over-time change in the ranking score (the gainscore). The use of Internet sites as a resource increased a small (not statistically significant) amount for students in the CS1 course, while LC students showed a drop in use. Conversely, use of professional organizations increased in the LC group while decreasing among students in the CS1 course.

**Table 4. Mean scoring for LC and CS 1 groups**

| Resource | CS 1 Pre | CS 1 Post | LC Pre | LC Post |
|---|---|---|---|---|
| Internet/web sites | 5.45 | 5.74 | 5.68 | 4.99 |
| Professional organizations | 2.54 | 2.26 | 2.60 | 3.34 |

Table 4 shows the mean scoring for CS1 and LC students before and after the course. Table 5 shows the change in response (the gainscore) for each group, as well as the result of the post-hoc t-test comparing the two. For this test, $H_0$ is that the mean gainscore of the two groups are equal.

**Table 5. Post-hoc test for differences LC vs. traditional**

| Resource | CS 1 Gainscore | LC Gainscore | Result (T-K adj. p) |
|---|---|---|---|
| Internet/web sites | 0.30 | -0.70 | t(320)=2.11, p=.0039 |
| Professional organizations | -0.29 | 0.74 | t(320)=-3.14, p=.0019 |

## 4. DISCUSSION

As mentioned previously, in this work we see the first statistically significant differences pre- to post-quarter among participants in the learning community as compared to students in the CS1 course.

### 4.1 Impact of the Learning Community

The learning community was shown to have a statistically significant impact over time in reducing the students' feeling of isolation in computer science courses. Students who took part in the learning community were more likely than their compatriots in the CS1 course to report a reduced feeling of isolation in the post-quarter survey than in the pre-quarter survey. This result confirms an effect seen in the analysis of the smaller learning community data sets [16] and is a sign that one of the purposes of the learning community was achieved during the first two cohorts.

The increased use of professional organizations as a resource was also seen among students in the learning community. Among the larger CS1 course population, no significant change occurred between the pre-quarter and post-quarter surveys. The learning community students started out with almost the same ranking assigned but left the course having reported a more frequent reliance on professional organizations. Some of this effect may be explained by the extra- and co-curricular activities organized for the learning community. Students visited several tech-focused companies as a part of the learning community, and this may have encouraged their utilization of professional organizations.

### 4.2 Limitations

There are some limitations to this study. Some of the questions posed in the survey are sensitive in nature, and participants may have chosen to not complete those portions of the survey causing their data to be discarded in the analysis. Students were asked to provide their student IDs as a part of the survey to enable verification that only those students who completed both parts were included in the analysis, and this may have led students to not participate in the survey. The evaluation of the learning community is based on self-reported values provided by participants, and care must be taken when interpreting the results. Additionally, though care was taken in choosing survey questions and choices that are unambiguous, there is a risk that the participant may have misinterpreted the questions or choices. Further the learning community population is relatively small even including both cohorts, making it difficult to determine any impact of the learning community on subpopulations.

## 5. CONCLUSION

Here we present the first results demonstrating an impact of participation in the learning community on student attitudes toward computing. We find a statistically significant reduction in reported isolation among learning-community students, a result that was not seen in students taking the programming course alone. We also see a statistically-significant increase in utilization of professional organizations among learning-community students.

## 6. ACKNOWLEDGEMENTS

Our thanks to Brian Sedlak for coding the data for the 2015 surveys.